\documentclass[conference]{IEEEtran}
\usepackage{cite}
\usepackage{newtxmath}
\usepackage{amsmath}
\usepackage{graphicx}
\usepackage{textcomp}
\usepackage{multirow}
\usepackage{xcolor}
\usepackage{hyperref}

\def\D{\phantom{0}}
\newcommand{\p}[1]{\phantom{#1}}

\usepackage{listings}
\definecolor{dkgreen}{rgb}{0,0.6,0}
\definecolor{gray}{rgb}{0.5,0.5,0.5}
\definecolor{mauve}{rgb}{0.58,0,0.82}
\usepackage{framed}
\newcommand{\Conclusion}[1]{\begin{framed}\noindent #1\end{framed}}

\begin{document}

\title{Characterising the Knowledge about Primitive Variables in Java Code Comments}

\author{
 \IEEEauthorblockN{Mahfouth Alghamdi}
 \IEEEauthorblockA{School of Computer Science\\
 The University of Adelaide\\
 Adelaide, Australia}
 \and
 \IEEEauthorblockN{Shinpei Hayashi}
 \IEEEauthorblockA{School of Computing\\
 Tokyo Institute of Technology\\
 Tokyo, Japan}
 \and
 \IEEEauthorblockN{Takashi Kobayashi}
 \IEEEauthorblockA{School of Computing\\
 Tokyo Institute of Technology\\ 
 Tokyo, Japan}
 \and
 \IEEEauthorblockN{Christoph Treude}
 \IEEEauthorblockA{School of Computer Science\\
 The University of Adelaide\\
 Adelaide, Australia}
 }
\maketitle

\begin{abstract}
Primitive types are fundamental components available in any programming language, which serve as the building blocks of data manipulation. Understanding the role of these types in source code is essential to write software. Little work has been conducted on how often these variables are documented in code comments and what types of knowledge the comments provide about variables of primitive types. In this paper, we present an approach for detecting primitive variables and their description in comments using lexical matching and advanced matching. We evaluate our approaches by comparing the lexical and advanced matching performance in terms of recall, precision, and F-score, against 600 manually annotated variables from a sample of GitHub projects. The performance of our advanced approach based on F-score was superior compared to lexical matching, 0.986 and 0.942, respectively. We then create a taxonomy of the types of knowledge contained in these comments about variables of primitive types. Our study showed that developers usually documented the variables' identifiers of a numeric data type with their purpose~(69.16\%) and concept~(72.75\%) more than the variables' identifiers of type String which were less documented with purpose~(61.14\%) and concept~(55.46\%). Our findings characterise the current state of the practice of documenting primitive variables and point at areas that are often not well documented, such as the meaning of boolean variables or the purpose of fields and local variables.  

\end{abstract}


\section{Introduction and Motivation}\label{sec:intro}

While modern source code hosting sites, such as GitHub, make a massive amount of source code available, reusing this code often requires reading and understanding code written by other developers. This task can be quite difficult, particularly when source code is not adequately commented. Commenting source code is a good programming practice as it can help developers to quickly understand source code for later modification, maintenance or reuse. For example, studies have shown that commenting source code can help improve the readability of source code~\cite{tenny1988program,tenny1985procedures}, while in other studies, researchers are of the view that code comments are an essential part of software maintenance~\cite{hartzman1993maintenance,jiang2006examining}. Code comments can provide additional information to help developers perform a wide range of software engineering tasks. For instance, code comments can be used for bug detection~\cite{rubio2010expect,silva2003importance,subramanian2014live}, specification inference~\cite{blasi2018translating,pandita2012inferring} and testing~\cite{goffi2016automatic,wong2015dase}.

In this paper, we investigate the role of primitive variable identifiers in comments, especially how commonly these identifiers are documented in accompanying comments and what type of additional information the comments contain about these variables. To lay the foundation for understanding the role of primitive variables and their documentation in Java source code, our preliminary findings show that the percentage of primitive data types in 2,491 Java repositories from GitHub was dominating with 60.10\%~(8,646,435), compared to non-primitive data types with 39.90\%~(5,741,380). 

In addition, while non-primitive types tend to have their intention encoded in the name of the type, the names of primitive types do not allow for the encoding of such knowledge, such as variable's purpose, concepts and directives. The introduction of Generics in Java has made even more type names self-explanatory~(e.g., List$<$File$>$ instead of List) while similar advantages are not available for primitive variables, and a developer often must guess what a String or float might contain. This motivated us to study the knowledge available about primitive variables in Java code comments. For instance, Listing~\ref{lst:code-example} shows a comment that is not informative with respect to the local variable ``ry''. Therefore, it will take developers quite some effort to understand the purpose, concepts and directives related to this identifier in the source code. This is due to 1) the developer did not explicitly mention the identifier in the comment nor followed the naming conventions to encode such types of knowledge in the variable’s identifier and 2) the lack of additional information about it in the comment. As a consequence, this can make program comprehension more difficult and ultimately lead to a reduction in productivity of the developer. Similar cases to the example shown in Listing~\ref{lst:code-example} are what we are targeting in our paper to identify knowledge types related to the variables' identifiers in their accompanying comments.

\begin{lstlisting}[label={lst:code-example},caption={Comment with respect to local variable ``ry''~\cite{Listing1}. 
},captionpos=b]
// CURVE-INSIDE
final float ry = t * (t * Ay + By);
\end{lstlisting}

In this work, we introduce the first empirical study to detect primitive variables’ identifiers in comments using two levels of matching techniques, to characterise the knowledge they contain. We contribute with the following:

\begin{itemize}
  \item We developed lexical and advanced matching techniques to capture the identifiers of primitive variables in Java source code comments, and then evaluated these approaches using a manually curated benchmark of six well-commented project repositories~\cite{appendix} hosted on GitHub.
  
  \item We manually classified the documented information, used to describe the variable identifiers in comments, into three types of knowledge: purpose, concept and directives.   

  \item A large-scale analysis of 2,491 engineered Java software repositories~\cite{appendix} hosted on GitHub was carried out to provide an insight into how developers document these variables in the form of source code comments.  
\end{itemize}

The remainder of this paper is structured as follows. We introduce our research questions in Section~\ref{sec:rereasch-questions}. Our detection techniques are described in detail in Section~\ref{sec:APPROACH}. In Section~\ref{sec:reserch-method}, we describe our study design as well as the methods used for data collection and analysis. Our findings are reported separately for each of our research questions in Section~\ref{sec:findings}. We then discuss threats to validity in Section~\ref{sec:thrests-validity} and related work in Section~\ref{sec:related-work}. Finally, we conclude the paper and outline future work in Section~\ref{sec:conclusion}.

\section{Research Questions}\label{sec:rereasch-questions}

Our ultimate goal of this paper is to reveal the nature of the documented information for variables in source code comments. We define documented variables as those variables that have comments above their declaration, and the developers mention the identifiers in their accompanying comments. To analyse the documented variables’ information, we had to capture code comments that include the variables’ information, using lexical matching. However, lexical matching~(i.e., exact matching) may not be considered the best choice to detect an identifier in a comment~(see Listing~\ref{example:lexical-vs-advanced}) because the variable’s identifier may not explicitly appear in the comments, e.g., developers use abbreviations for the identifier or explain it with different terms. To tackle such a problem, we developed a detection method considering lexical and advanced matching, which has the ability to capture the exact identifier and its meaning in a comment.

\begin{lstlisting}[caption={Example to show the inability of lexical matching to detect variables' identifiers in a comment~\cite{Listing2}.},%
                   label={example:lexical-vs-advanced},captionpos=b]
/**
Writes a message into the database table.
Throws an exception, if an database-error occurs !
*/
public void append(String _msg) throws Exception {
\end{lstlisting}

According to the best of our knowledge, there is no guideline to identify a variable’s identifier in a comment, although researchers have used different techniques in their work~\cite{chen2019automatically}. Therefore, we proposed and evaluated two detection techniques for detecting the variables’ identifiers in comments, with a manually annotated benchmark of six well-commented project repositories hosted on GitHub, and we then investigate the nature of the documentation for variables. The creation of the manual evaluation data set was necessary to evaluate our approaches, as a reference data set does not exist in the literature. To assess the quality of our detection approaches, we ask our first research question.

\begin{itemize}
  \item \textbf{RQ1}: To what extent can different techniques detect variables in comments? \begin{itemize}
    \item \textbf{RQ1.1}: To what extent is the lexical matching technique able to detect the variable names in the comments?
    \item \textbf{RQ1.2}: To what extent are different advanced matching techniques able to detect the meaning of variable names in the comments? 
  \end{itemize}
\end{itemize}

The answer of RQ1 would confirm the quality of our detection approaches to be used for large-scale analyses, to answer RQ3--5.

As the identifiers of primitive variables require developers to provide additional information to explain the functionality of the variables in the comments, we further ask the following question: 

\begin{itemize}
  \item \textbf{RQ2}: What types of knowledge do comments provide about the variables? 
\end{itemize}

The method used to answer RQ2 was inspired by previous work~\cite{maalej2013patterns}. As far as we know, the types of knowledge relevant to variables have not yet been studied in the literature. To this aim, we identify three knowledge types that are closely related to the variable's domain significance and based on previous work to fit our propose to answer this question.  

Since there is little known about the prevalence of primitive variables and their documentation used by developers, we aim to investigate how these variables are used and documented in source code comments by asking the following questions. 

\begin{itemize}
  \item \textbf{RQ3}: How frequently are primitive variables documented in comments?

  \item \textbf{RQ4}: What are the distributions of documented variables by their scope of declarations?

  \item \textbf{RQ5}: What are the types of comments associated with the scopes of the documented variables? 
\end{itemize}

\section{Detecting Documented Variables}\label{sec:APPROACH}

In this section, we present our techniques to detect variables’ identifiers in comments. To this end, several detection approaches were developed, using lexical matching, advanced matching and the union of lexical and advanced matching.

\subsection{Preprocessing}\label{subsec:preprocessing}
Preprocessing variables’ identifiers and comments is an essential step for our detection approaches. For this, we followed part of the steps proposed by Ratol et al.~\cite{ratol2017detecting} for preprocessing both variables’ identifiers and the text of comments, while we create our own prepossessing steps to suit our matching techniques. We present these steps as following: 

\textbf{Identifiers.} We split an identifier based on its typographical conventions to either a single term or multiple terms~(i.e., compound terms) using the camelCase rules and other rules based on regular expressions we designed, such as splitting the identifier using underscore.  
We then convert each term to its root using the Lemmatiser of the Stanford Core NLP library~\cite{manning2014stanford} while we maintain the original term for later use. As a consequence, each term produced a dictionary entry containing $\langle\text{term},\text{lemma}\rangle$.

\textbf{Comments.} We processed the comments by first splitting them into sentences, using the sentence detector model in the OpenNLP library. 
Each sentence was then further split into tokens using the same library. Tokens that are detected to be compound terms are further split into one or more terms in the same way the identifiers were split. After tokenisation, we remove stop words such as ``the'' and ``or'' from the list of tokens obtained, unless these stop words are part of the variables’ identifier. Finally, we generate a dictionary containing $\langle\text{token},\text{lemma}\rangle$ for each token in the list.

\textbf{Implementation note.} 
The lemmatisation is only applied for the variables’ identifiers and comments’ tokens, for lemma and metaphonic advanced matching techniques.

\subsection{Lexical Matching}
Our approach using lexical matching is aimed at searching for exact~(i.e., literal) matching of variables’ identifiers in the head comments. For example, in Listing~\ref{example:lexical-matching}, lexical matching can detect the variable identifier \emph{ignore} in the provided comment. Lexical matching may suffer from different issues, such as spelling and typographical errors, which may hinder its detecting performance, and yet it can show how often developers document the variable’s identifier in comments with its exact form.

\begin{lstlisting}[caption={Example of matching using lexical technique~\cite{Listing3}.},%
                   label={example:lexical-matching},captionpos=b]
//if ignore is true, this column will be ignored by building sql-statements.
boolean ignore = false;
\end{lstlisting}

\subsection{Advanced Matching}
To tackle the problem indicated in lexical matching, we introduced a new way of detection, i.e., the meaning of the variable’s identifier can be captured through a set of words in the comment. Our advanced matching approach is composed of lemmas’ matching, metaphonic matching and SEthesaurus matching.

\subsubsection{Lemmas Matching}
Lemmatisation is the process of grouping together the different inflected forms of a word into a single term. Our lemmas’ matching approach works by lemmatising the variable’s identifier and the words in the text of the comment. It then looks for each identical lemma term that can exist between the identifier and the token in the text of a comment. If all lemmas of the identifier are found, it can be said that the variable was documented in the comments. For instance, in Listing~\ref{example:advanced-matching-lemmas}, the comment contains two lemma terms, matching the two lemma terms of the identifier.

\begin{lstlisting}[caption={Example of advanced matching using Lemmas technique~\cite{Listing4}.},%
                   label={example:advanced-matching-lemmas},captionpos=b]
// Make sure at least one connection for this protocol succeeds (if expected to)
boolean connectionSucceeded = false;
\end{lstlisting}

\subsubsection{Metaphonic Matching}
Our second advanced detection approach is based on the Double Metaphone encoding algorithm~\cite{philips2000double}. The algorithm is a searching technique that can handle different matching problems due to misspelling of given keywords. It works by taking two input strings and returns true if they phonetically match---the algorithm takes into its consideration the similar sounds produced by different characters. For example, the encoding of a misspelled word will often match the encoding of the word that was intended. For example, in Listing~\ref{example:advanced-matching-metaphonic}, the Double Metaphone algorithm encodes the identifier \emph{configured} and the term \emph{configuration} which appears in the comment with code of "KNFK" because they phonetically match. However, using the same example for the detection of the variable name in the comment using the lemma and SEthesaurus matching techniques will yield no matches.

\begin{lstlisting}[caption={Example of advanced matching using Metaphonic technique~\cite{Listing5}.},%
                   label={example:advanced-matching-metaphonic},captionpos=b]
//A flag to indicate configuration status
private boolean configured = false;
\end{lstlisting}

\subsubsection{SEthesaurus Matching}
The last matching technique used for the detection of the meaning of variable’s identifier in the head comment is based on SEthesaurus~\cite{chen2017synonym}. SEthesaurus covers a large set of software-specific terms~(52,645 terms), counting 4,773 abbreviations and 14,006 synonym groups, with high accuracy; this can be beneficial, in our case, to detect such abbreviations and the synonyms between the variables’ identifiers and terms in the text of the comments. For instance, in Listing~\ref{example: advanced-matching-SEthesaurus}, the variable’s identifier \textit{TRACE\_INT} consists of two terms: 1) TRACE and 2): INT, where the second term of the identifier \emph{INT} is an abbreviation of its full form \emph{integer}, which appears in the comment. It is worth noting that lemmas’ and metaphonic matching can fail to detect the variables’ identifiers in the text of the comments in similar situations.

\begin{lstlisting}[caption={Example of advanced matching using SEthesaurus technique~\cite{Listing6}.},%
                   label={example: advanced-matching-SEthesaurus},captionpos=b]
/**
 * TRACE level integer value.
 * 
 * @since 1.2.12
 */
public static final int TRACE_INT = 5000;
\end{lstlisting}

\subsection{Union of Matching Approaches}
As each detection approach has its own ability to detect the variable’s identifier in the comments, we considered using the union of lexical and advanced matching approaches to overcome performance limitations and provide high detection coverage of the identifiers in the comments. We considered using two ways to unify these approaches: 1) a union of the three advanced approaches~(i.e., lemmas’, metaphone and SEthesaurus) and 2) a union of all the approaches~(i.e., lexical and advanced approaches).

\section{Study Design}\label{sec:reserch-method}

\subsection{Data Collection}\label{subsec:data-collection}

This section outlines the data set used to evaluate our approaches, the data set used for large-scale analysis, and the types of variables, comments, and scopes of the variables' declarations. 

\subsubsection{Repositories}
Our repository preparation considers the data set used to evaluate our detection approaches and to study common characteristics of the documentation of primitive variables using large-scale quantitative analysis.
\begin{figure}[tb]\centering
  \includegraphics[width=\linewidth]{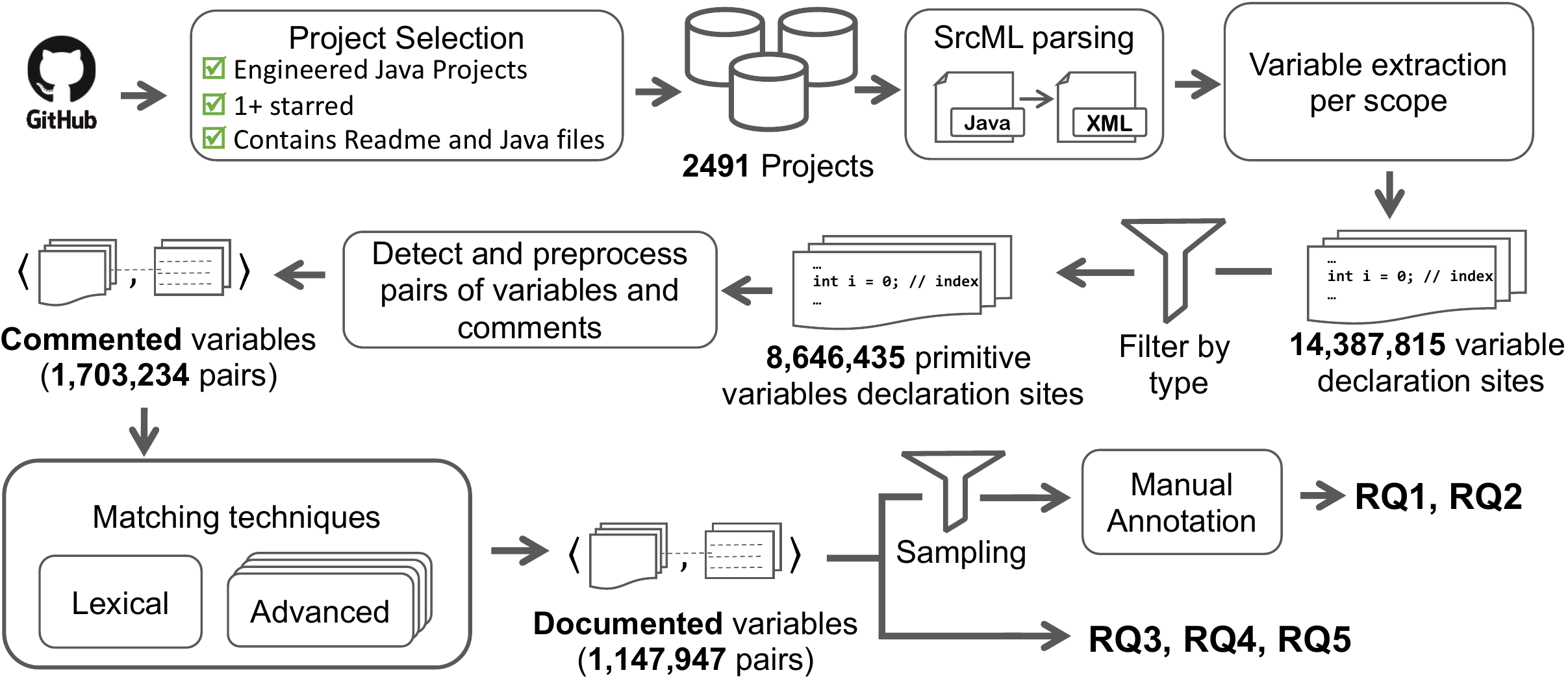}
  \caption{Overview of our study.}\label{fig:approach}
\end{figure}

\begin{table}[tb]\centering
\caption{Used Variables for Our Detection Approaches}\label{tab:dataset-evaluation-and-large-scale-analysis-haveComment}
{\tabcolsep=5pt\begin{tabular}{l|rr|rr}\hline
\textbf{Data} & \multicolumn{2}{c|}{\textbf{Evaluation data set}} & \multicolumn{2}{c}{\textbf{Large-scale data set}} \\
\textbf{types} & \textbf{\# vars} & \textbf{\# commented} & \textbf{\# vars} & \textbf{\# commented} \\ \hline
boolean & 1,161 & 261 \D(4.91\%) & 733,294 & 125,598 (17.13\%) \\
byte & 2,755 & 181 (11.65\%) & 187,432 & 32,762 (17.48\%) \\
char & 158 & 29 \D(0.67\%) & 54,116 & 9,833 (18.17\%) \\
int & 7,592 & 1,952 (32.12\%) & 2,505,266 & 424,265 (16.93\%) \\
short & 58 & 12 \D(0.25\%) & 30,703 & 6,792 (22.12\%) \\
long & 4,715 & 976 (19.95\%) & 640,504 & 83,088 (12.97\%) \\
float & 76 & 26 \D(0.32\%) & 134,263 & 26,868 (20.01\%) \\
double & 669 & 160 \D(2.83\%) & 291,384 & 56,707 (19.46\%) \\
String & 6,456 & 1,501 (27.31\%) & 4,069,473 & 937,321 (23.03\%) \\ \hline
Total & 23,640 & 5,098 (21.57\%) & 8,646,435 & 1,703,234 (19.70\%) \\ \hline
\end{tabular}}%
\end{table}

\begin{table}[t]\centering
\caption{Amount of Primitive and Other Variables}\label{tab:primitivesAndOtherVariables}
\begin{tabular}{lll}\hline
                 & \textbf{Primitive variables} & \textbf{Other variables} \\ \hline
 Fields          & 1,697,033        &\p{0,}537,067 \\
 Local variables & 3,020,165           & 3,693,085 \\
 Parameters      & 3,929,237           & 1,511,228 \\ \hline
 Total           & 8,646,435 (60.1\%)  & 5,741,380 (39.9\%)  \\ \hline
\end{tabular}%
\end{table}

\begin{table*}[tb]\centering
\caption{Evaluation Sample}\label{tab:evaluation-sample} 
\begin{tabular}{ll|rrr|rrr|rrr|rr} \hline
\multicolumn{2}{c|}{\textbf{Projects}} & \multicolumn{3}{c|}{\textbf{Fields}} & \multicolumn{3}{c|}{\textbf{Local variables}} & \multicolumn{3}{c|}{\textbf{Parameters}} & \multicolumn{2}{c}{\textbf{Total}} \\
\textbf{Name} & \textbf{Version} & \textbf{Total} & \textbf{Pop.} & \textbf{Smp.} & \textbf{Total} & \textbf{Pop.} & \textbf{Smp.} & \textbf{Total} & \textbf{Pop.} & \textbf{Smp.} & \textbf{Pop.} & \textbf{Smp.} \\ \hline
Chronicle Map & 3.19.42 & 452 & 10 & 7 & 1,359 & 16 & 12 & 1,494 & 111 & 81 & 137 & 100 \\
Joda time & 2.10.6 & 847 & 82 & 5 & 2,230 & 26 & 2 & 3,559 & 1,392 & 93 & 1,500 & 100 \\
JUnit & 4.13 & 292 & 1 & 0 & 320 & 0 & 0 & 721 & 201 & 100 & 202 & 100 \\
Log4j & 1.9.0 & 813 & 126 & 28 & 1,053 & 12 & 2 & 1,330 & 313 & 70 & 451 & 100 \\
Spring Data Redis & 1.2.18 & 539 & 5 & 1 & 1,076 & 2 & 0 & 4,910 & 924 & 99 & 931 & 100 \\
JFlex & 2.3.1 & 755 & 109 & 27 & 989 & 31 & 7 & 901 & 269 & 66 & 409 & 100 \\ \hline
Total & --- &3,698 & 333 & 68 & 7,027 & 87 & 23 & 12,195 & 3,210 & 509 & 3,630 & 600\\ \hline
\end{tabular}%
\end{table*}

\textbf{Dataset used for large-scale analysis.}
Motivated by previous work on the promises and perils of mining GitHub~\cite{kalliamvakou2014promises}, which concluded that many repositories on GitHub do not contain engineered software projects, we used the RepoReaper framework~\cite{munaiah2017curating} to obtain repositories for this work. RepoReaper was developed to differentiate between repositories with engineered software projects and those with noise~(e.g., assignments projects). This is an important step, as noise projects could be the cause of incorrect conclusions in our study.

To select repositories for our study, we first obtained 17,243 Java projects, which had at least one star and were classified as containing engineered software projects by the Random Forest classification of RepoReaper. We then eliminated 680 projects that did not exist anymore on GitHub~(e.g., deleted or made private). For the remaining projects~(16,563), we filtered out 1,441 projects that did not have README files and those projects~(12,578) that did not have the word \emph{documentation} in their README files—--this step was necessary as we wanted to obtain only well-documented projects. From the remaining projects~(2,544), we removed 53 projects that did not contain any Java files. Finally, in total, 1,040,026 Java files were collected from 2,491 projects, which were used in this study.

\textbf{Dataset used for evaluation.}
Six project repositories were used to evaluate our proposed approaches, motivated by previous work of Ratol et al.~\cite{ratol2017detecting} to detect fragile comments. These projects~(see Table~\ref{tab:evaluation-sample}) were selected because they were used in Ratol et al.'s work~\cite{ratol2017detecting} and showed several advantages such as their availability as open-source, a diversity of application domains, and being well-commented. To allow further investigation, the six projects were cloned locally to ease the process to analysing them.

The same work by Ratol et al.~\cite{ratol2017detecting} was followed to determine our sample size~(i.e., we utilized the stratified random sampling strategy to achieve diversity of the variables in our sample). Stratified random sampling can prevent bias while ensuring that all classes of interest are covered in a sampled population. Following the procedure in this strategy, we randomly selected 100 variables from each of the six projects, in proportion to the number of variables declared in each scope~(i.e., fields, local variables and methods’ parameters) from the target population---we defined target population as the primitive variables that have at least one comment above their declaration, and the variables' identifier appeared in the comment. For each project in Table~\ref{tab:evaluation-sample}, we provide the total number of variables in a given scope, followed by the number of variables~(in column \emph{Pop}) of this scope for which our approach can detect the variable name in the comment above those variables’ declarations. In column \emph{Smp}, we indicate the number of variables in this scope, which were selected for the sample. For example, for the project \emph{Chronicle Map}, the total number of variables detected in the field scope is 452, and there are 10 variables for which our approach can find a match of these variable names in the comments above their declaration. Finally, the total number of variables’ identifiers that our advanced and lexical approaches detected with associated comments in the Chronicle Map project is 137.

\subsubsection{Types of the Variables, Comments, and Scopes}
The source code in any programming language is divided into small pieces of code elements, e.g., classes, methods, fields, etc. According to the syntax of the Java programming language, there are four types of comments: 1) inline comment~(i.e., \texttt{//} shown in the same line where the variable is declared), 2) line comment~(i.e., \texttt{//} shown above variables' declarations), 3) block comment~(i.e., \texttt{/*} $\cdots$ \texttt{*/}), and 4) Javadoc comment~(i.e., \texttt{/**} $\cdots$ \texttt{*/}). 

In this work, we focus on detecting the identifiers of primitive variables: boolean, byte, char, int, long, float, double and String that are documented in these three types of comments~(i.e., line and inline, block and Javadoc), declared in three categories of program elements~(i.e., fields, local variables found in methods’ bodies and parameters of the methods). We considered String as a primitive data type in Java source code, motivated by the fact that it is a primitive type in other languages, such as Python. As the first step, we collected all variables~(i.e., 1,703,234) that have at least one comment above their declarations~(Table~\ref{tab:dataset-evaluation-and-large-scale-analysis-haveComment}), and we then used our detection approaches to identify these identifiers in their associated comments.

\subsubsection{Detecting Documented Variables in Source Code} We developed two approaches and then combined these two approaches to detect the variables’ identifiers in the source code comments. These approaches are based on lexical or advanced matching~(see Section~\ref{sec:APPROACH}) to capture variables’ identifiers in their accompanying comments. In lexical matching, our matching technique captures exact matches~(i.e., case sensitive) of these identifiers in the comments, while advanced matching captures the meaning of the identifier in the comments by applying lemmas, metaphonic and SEthesaurus  matching techniques.

\begin{table*}[tb]\centering
\caption{Annotation Questions}\label{tab:annotation-questions}
\begin{tabular}{lp{5cm}cp{8cm}cp{3cm}} \hline
& \textbf{Question} & \textbf{Answer} & \textbf{Description} & \textbf{avg.~$\kappa$} \\ \hline
AQ1 & Was the variable name mentioned in the comment? & Yes / No & N/A & 0.93\\ \hline
AQ2 & Did the comment add additional information other than just mentioned the variable name? & Yes / No & N/A & 0.82 \\ \hline
AQ2.1 & Did the comment provide any additional knowledge type about the purpose or patterns of the variable? & Yes / No & Explanation about the lifecycle of the variable in the comment; how to be referred/defined. This might be a part of the post-conditions of a method. In addition, the description of the initialisation process of the field variable can be treated as this category. & 0.72 \\ \hline
AQ2.2 & Did the comment provide any additional knowledge types about the concept of the variable? & Yes / No & Explanation about the content of the variable in the comment. Sometimes, the variable's identifier is enough to be explained. & 0.72 \\ \hline
AQ2.3 & Did the comment provide any additional knowledge types about the directive of the variable? & Yes / No & Explanation about the variable's domain, type or nullability as a part of the pre-requisite for a method. & 0.85 \\ \hline
AQ3 & Was the variable meaning mentioned in the comment? & Yes / No & Explanation about the variable's identifier in term of the meaning, which used to describe the variable in the comment. & 0.75 \\ \hline
\end{tabular}%
\end{table*}

Our approach to detect the documented variables in their associated comments is shown in Figure~\ref{fig:approach}. It is worth noting that projects used in our input source are classified as engineered software projects based on the RepoReaper tool~\cite{munaiah2017curating}, which was developed to distinguish between engineered projects and other projects. We make use of RepoReaper's Random Forest classification, trained with organisation and utility data sets.

After cloning the targeted projects from GitHub and removing all non-Java files, we made use of srcML~\cite{collard2011lightweight} to convert the source code of each .java file into its XML representation. The advantage of using srcML is its ability to perfectly preserve the format of the original source at different levels, such as lexical, documentary~(e.g., comments, white space), structural~(e.g., classes, methods) and syntactic~(e.g., statement).

Each of the XML versions was then used to facilitate the detection and the extraction of the variables with their head comments~(i.e., comments that appeared immediately above the variables’ declaration) and their scopes in which they appeared. Note: the white-spacing between the variables’ declarations and their head comments were ignored. Finally, out of 14,387,815 variables obtained from 2,491 projects, we extracted 60.10\%~(8,646,435) as primitive variables~(see Table~\ref{tab:primitivesAndOtherVariables}). 19.70\%~(1,703,234) of these primitive variables had comments above their declarations, as shown in Table~\ref{tab:dataset-evaluation-and-large-scale-analysis-haveComment}, to be used as input sources for our detection approaches.


\subsection{Data Analysis}\label{data-analysis}
In this section, we outline the data analysis methods used to answer the research questions.

\subsubsection{Accuracy of Detection Approaches}
To evaluate the accuracy of our lexical and advanced matching approaches, four authors of this study manually annotated 600 variables to investigate the presence of the variables’ identifiers in the text of the comments. We considered the union of these techniques~(lexical and advanced matching) to capture the variables in comments, and we evaluated the accuracy of each of these approaches individually. To calculate inter-rater agreements between the annotators, the annotators were split into six pairs~(i.e., each of the four authors was paired with each of the other authors) where each pair annotated 100 randomly selected variables. Each pair of annotators proceeded to annotate each of the identifiers with their related comments to evaluate the accuracy of the matching techniques, and the presence of the additional information~(see Table~\ref{tab:annotation-questions}). We then measured the agreement reliability between pairs of annotators using Cohen's Kappa metric. We note that the kappa reported is the average kappa value across all annotator pairs for each annotation question, which shows substantial to almost perfect agreement~\cite{landis1977measurement}. We further studied these disagreements between the pairs of annotators and resolved all conflicts to enhance the kappa agreement values for each question, i.e., almost perfect agreement.

Table~\ref{tab:annotation-questions} shows our annotation questions, the answer options for each one, a description that shows how to answer each of these questions, and the average kappa value across all the annotators' pairs. Each pair of annotators used AQ1 and AQ3 to evaluate the lexical and advanced matching techniques’ performance, respectively. AQ1 investigates whether a variable identifier is literally documented in the comment. In contrast, AQ3 investigates whether the meaning of the identifier is documented in the comment. Finally, AQ2 investigates the existence of any type of knowledge regarding purpose, concept and directives found in the comment, associated with the identifiers and based on each answer to the sub-questions.

To annotate a variable’s identifier based on AQ1, which is used to evaluate the performance of the lexical matching, we differentiated between variable identifiers by their typographical conventions; for example, an identifier consists of a single term or multiple terms as described in Section~\ref{subsec:preprocessing}. In case of an identifier composed of multiple terms, each annotator searches for all terms in the identifier and the corresponding words with the exact wording of each term in the text of the comment. The manual annotation steps consider using the same sequence for the comment in which the terms appeared in the identifier while white-spacing between these terms was ignored. On the other hand, in case of an identifier containing only a single term, the annotator looks at the literal matches between the identifier and its corresponding word in the comment.

The performance of each of our matching techniques is then separately measured using the commonly used evaluation measures: recall, precision, and the F-score. Recall measures the degree of absence of false negatives while precision measures the degree of absence of false positives. For example, perfect recall reported by one of the approaches can indicate that such a technique was able to detect all the variables’ identifiers in their comments, which were also detected by the manual inspection. It is worth mentioning that the ``recall'' reported in this paper is an approximation, as a theoretical value of the recall cannot be computed precisely because much effort to manually annotate the data would be required.

\subsubsection{Knowledge Types Present in Comments}
Since there is no existing taxonomy of information in the comments which would provide additional information around the variable identifiers, we established three types of knowledge, inspired by previous work of Maalej and Robillard~\cite{maalej2013patterns}. The description of these types was only applicable to API documentation. However, we identified three types of knowledge from their work which was applicable in our context with some adjustments, namely purpose, concept and directives, and we then provided a description for each one of those types to fit our purpose, as illustrated in AQ2.1--2.3 in Table~\ref{tab:annotation-questions}. 

Each pair of annotators then inspected each identifier and its accompanying comment to determine the existence of identifiers’ purpose, concept and directives. We then quantitatively analysed how many of the variables’ identifiers were documented in their accompanying comments with one or more of these types of knowledge.

\subsubsection{Number of Times Variables' Identifiers Commented in Source Code}
To investigate how frequently primitive variables are commented in the source code, we first obtained all the variables that have at least one comment~(1,703,234 variables). Then, we quantitatively analysed the variables documented using the best matching approach, resulting from RQ1, to determine how many of the variables’ identifiers are mentioned in their comments in different scopes.

\subsubsection{Distribution of Documented Variables by Their Scope of Declarations}
Our investigation in this study takes into account the type of documented identifier and the scope in which they are declared. We hypothesise that some data types of variables tend to be documented in the comments in particular scopes more than other data types. Thus, to explore the distribution of these identifiers, we quantitatively analysed these variables by their scopes of declarations.

\subsubsection{Types of Comments Associated with the Documented Variables by Scope}
We further investigated the types of comments associated with the documented data types of variables. For example, we investigated the scopes of the variables in which they are declared, the data types of these variables and the types of the comments associated with these variables. We then analysed the descriptive statistics of our data set.

\section{Findings}\label{sec:findings}

In this section, we present our findings individually for each of the research questions. 

\subsection{RQ1: To what extent can different techniques detect variables in comments?}

\begin{table}[tb]\centering
\caption{Accuracy of Different Matching Techniques}\label{tab:RQ1-result}
{\tabcolsep=5pt\begin{tabular}{lrrr}\hline
\textbf{Matching techniques} & \textbf{Recall} & \textbf{Precision} & \textbf{F-score} \\ \hline
Lexical Matching & 0.896 & 0.994 & 0.942 \\
Advanced Matching (Lemmas) & 0.985 & 0.985 & 0.985 \\
Advanced Matching (Metaphone) & 0.995 & 0.973 & 0.984 \\
Advanced Matching (SEthesaurus) & 0.462 & 0.985 & 0.629 \\
Union of all advanced matchings & 1.000 & 0.973 & 0.986 \\ 
Union of lexical and advanced matching & 1.000 & 0.973 & 0.986 \\ \hline
\end{tabular}}%
\end{table}
To understand the performance of lexical and advanced techniques used to detect the variables’ identifiers in the comments, we present our results of the evaluation of the lexical and the advanced matching in Table~\ref{tab:RQ1-result}. The evaluation of these techniques was based on 600 variables manually annotated, as shown in Table~\ref{tab:evaluation-sample}, and using the annotation questions~(i.e., AQ3), shown in Table~\ref{tab:annotation-questions}, to assess the performance of all techniques. As a result of the evaluation process, we found that 584 variables’ identifiers were correctly detected by our approaches along with their accompanying comments, whereas only 16 variables were not detected correctly.

The overall performance based on the F-score for each of the techniques used to detect variables’ identifiers, across all the scopes, showed that the ability of lemmas’ and metaphonic matching techniques to capture the variables’ identifiers in the comments were superior compared to the performances of the lexical and SEthesaurus approaches, i.e., the F-score measures for lexical, lemmas’, metaphonic and SEthesaurus are 0.942, 0.985, 0.984, and 0.629, respectively.

In terms of recall, SEthesaurus scored very low as compared to other approaches, and this can be attributed to the limited vocabulary in its dictionary, which was used to capture the abbreviations and synonyms of the identifiers in the text of comments. On the other hand, the recall of metaphonic matching, which can catch the spelling and pronunciation of the identifiers in the comments, scored the highest value among the approaches.

Furthermore, combining lexical and advanced matching resulted in a slight improvement~(4.3\%) of the F-score value, due to the few false positives instances being detected when the metaphonic matching technique was used.

The main observation we can draw from these results is that the F-score value of advanced matching techniques to detect the identifiers in the comments is higher than lexical matching. This means that advanced matching can detect the identifiers with 5\% more true positive instances than lexical matching~(555 true positive instances), which has ultimately impacted the value of the F-score of the union approach.

\Conclusion{%
The ability of the advanced matching to detect the variables’ identifier in comments based on F-score scored a higher value~(0.986), compared to lexical matching, which scored 0.942.}

 
\subsection{RQ2: What types of knowledge do comments provide about the variables?}

\begin{table}[tb]\centering
\caption{Relationship between Groups of Data Types of Variables with respect to Knowledge Types}\label{tab:PCD-Chi-Squire}
\begin{tabular}{clccc}\hline
 & \multirow{2}{*}{\textbf{Data type of variables}} & \multicolumn{3}{c}{\textbf{Types of knowledge}} \\
 & & \textbf{P} & \textbf{C} & \textbf{D} \\ \hline
\multirow{2}{*}{1} & Numeric & \multirow{2}{*}{0.0484} & \multirow{2}{*}{0.0001} & \multirow{2}{*}{0.0001} \\
 & String and char &  &  &  \\ \hline
\multirow{2}{*}{2} & Numeric & \multirow{2}{*}{0.3846} & \multirow{2}{*}{0.0067} & \multirow{2}{*}{0.0081} \\
 & boolean &  &  &  \\ \hline
\multirow{2}{*}{3} & String and char & \multirow{2}{*}{0.9053} & \multirow{2}{*}{0.6413} & \multirow{2}{*}{0.0001} \\
 & boolean &  &  &  \\ \hline
\end{tabular}%
\end{table}
 \begin{figure}[tb]\centering
  \includegraphics[width=\linewidth]{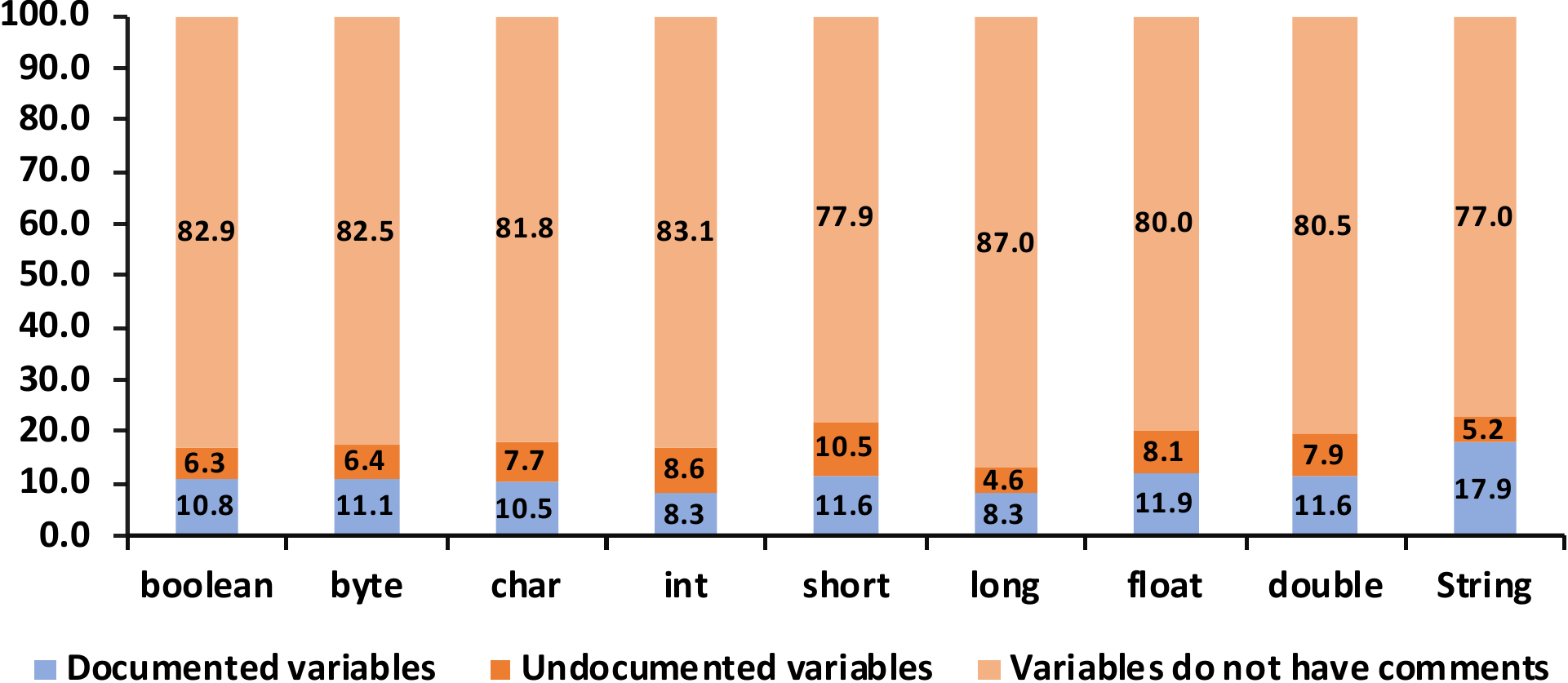}
  \caption{Percentage of variables documented, undocumented in the comment, and the percentage of the variables that do not have comments.}\label{fig:documnted-notDocumnted-notCommented-variables}
\end{figure}
\begin{figure}[tb]\centering
  \includegraphics[width=\linewidth]{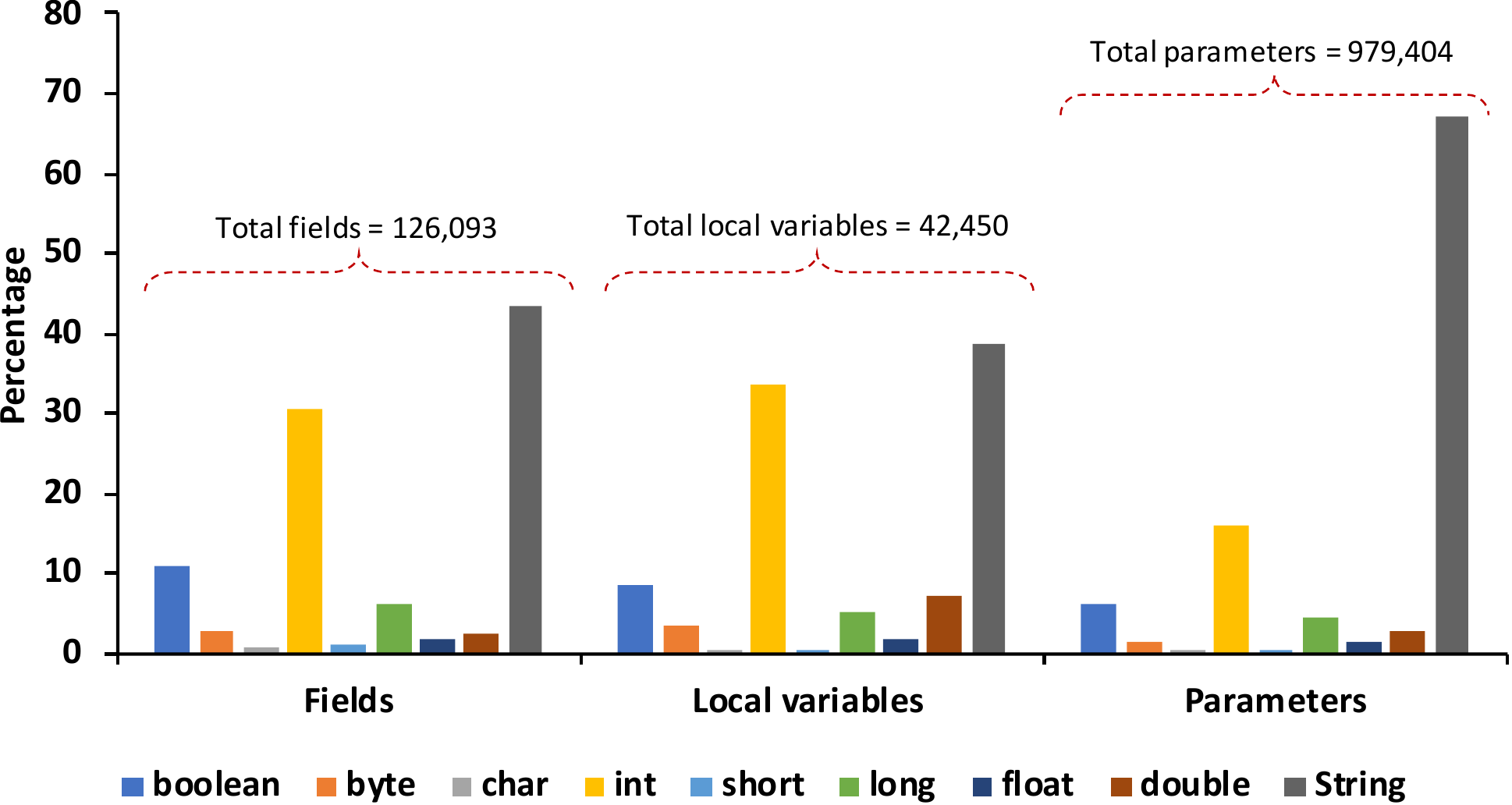}
  \caption{Percentage of variable data types declared in each scope.}\label{fig:data-types-per-scopes}
\end{figure}

\begin{table*}\centering
\caption{Types of Knowledge of Purpose (P), Concept (C), and Directives (D) Documented with Each Data Type of Variables in Comments}\label{tab:RQ2-result}
{\tabcolsep=3.5pt\begin{tabular}{l|rrrr|rrrr|rrrr|rrrr}\hline
\multirow{2}{*}{\textbf{Data type}} & \multicolumn{4}{c|}{\textbf{Fields}} & \multicolumn{4}{c|}{\textbf{Local variables}} & \multicolumn{4}{c|}{\textbf{Parameters}} & \multicolumn{4}{c}{\textbf{Combined}} \\
 & \textbf{Count} & \textbf{P} {\scriptsize (\%)} & \textbf{C} {\scriptsize (\%)} & \textbf{D} {\scriptsize (\%)} & \textbf{Count} & \textbf{P} {\scriptsize (\%)} & \textbf{C} {\scriptsize (\%)} & \textbf{D} {\scriptsize (\%)} & \textbf{Count} & \textbf{P} {\scriptsize (\%)} & \textbf{C} {\scriptsize (\%)} & \textbf{D} {\scriptsize (\%)} & \textbf{Count} & \textbf{P} {\scriptsize (\%)} & \textbf{C} {\scriptsize (\%)} & \textbf{D} {\scriptsize (\%)} \\ \hline
boolean & 10 & 50.00 & 60.00 & 0.00 & 1 & 100.00 & 100.00 & 0.00 & 26 & 65.38 & 46.15 & 0.00 & 37 & 62.16 & 51.35 & 0.00 \\
byte & 0 & N/A & N/A & N/A & 0 & N/A & N/A & N/A & 17 & 64.71 & 58.82 & 47.06 & 17 & 64.71 & 58.82 & 47.06 \\
char & 1 & 0.00 & 0.00 & 100.00 & 0 & N/A & N/A & N/A & 1 & 100.00 & 100.00 & 0.00 & 2 & 50.00 & 50.00 & 50.00 \\
int & 25 & 28.00 & 72.00 & 4.00 & 14 & 42.86 & 42.86 & 0.00 & 134 & 74.63 & 74.63 & 19.40 & 173 & 65.32 & 71.68 & 15.61 \\
short & 0 & N/A & N/A & N/A & 0 & N/A & N/A & N/A & 1 & 100.00 & 100.00 & 0.00 & 1 & 100.00 & 100.00 & 0.00 \\
long & 6 & 33.33 & 83.33 & 0.00 & 3 & 33.33 & 33.33 & 0.00 & 88 & 80.68 & 72.73 & 17.05 & 97 & 76.29 & 72.16 & 15.46 \\
float & 0 & N/A & N/A & N/A & 0 & N/A & N/A & N/A & 10 & 100.00 & 80.00 & 10 & 10 & 100.00 & 80.00 & 10.00 \\
double & 1 & 0.00 & 100.00 & 0.00 & 3 & 0.00 & 100.00 & 0.00 & 32 & 68.75 & 81.25 & 9.38 & 36 & 61.11 & 83.33 & 8.33 \\
String & 26 & 34.62 & 88.46 & 11.54 & 2 & 50.00 & 50.00 & 0.00 & 199 & 64.82 & 51.26 & 33.17 & 227 & 61.23 & 55.51 & 30.40 \\ \hline
Numeric & 32 & 28.13 & 75.00 & 3.13 & 20 & 35.00 & 50.00 & 0.00 & 282 & 76.24 & 74.11 & 18.79 & 334 & 69.16 & 72.75 & 16.17 \\
String and char & 27 & 33.33 & 85.19 & 14.81 & 2 & 50.00 & 50.00 & 0.00 & 200 & 65.00 & 51.50 & 33.00 & 229 & 61.14 & 55.46 & 30.57 \\ \hline
Total & 69 & 33.33 & 76.81 & 7.25 & 23 & 39.13 & 52.17 & 0.00 & 508 & 71.26 & 63.78 & 23.43 & 600 & 65.67 & 64.83 & 20.67 \\ \hline 
\end{tabular}}
\end{table*} 
The answer to RQ2 reveals the types of additional information in comments about documented variables, found in the 600 variables manually annotated. Table~\ref{tab:RQ2-result} shows three types of knowledge---purpose, concept and directive---that the developer might use to describe the identifiers in the comment. As shown in the Figure, the documented variables were grouped in scopes, and it shows the number of documented variables with any of these types of knowledge and percentages of the additional information of these knowledge types captured in the comment to describe the identifier.

For instance, our result revealed that 60.00\%~(6/10) of the boolean variables, declared in the field scope, have information that can explain the variables’ concept in their accompanying comments. On the other hand, 88.46\%~(23/26) of the String variables, declared in field scope, are documented with their concept. We can also observe that all variables declared in the parameters of the methods were documented with additional information of the three knowledge types. This indicates that parameter variables are crucial in source code, so developers tend to provide additional information about their concept, purpose or directives in their accompanying comments.

In addition, combining the numbers of each data type of variables, from all the scopes, can show how often variables with a particular data type were documented in comments with their concept, purpose or directives. In general, we can see only 20.67\%~(124) of these variables are documented in comments with their directive information, compared to the variables documented with their purposes 65.67\%~(394) and concept 64.83\%~(389)---developers rarely documented primitive variables with their directive information in the comments. 

We further grouped the data types of the primitive variables into numeric~(i.e., byte, int, long, short, float and double), String and char, and booleans to investigate whether data types of variables’ identifiers would affect the amount of the additional information regarding purpose, concept and directives associated with variables documented in comments. We found that only 55.46\%~(127/229) of the identifiers of strings and chars were documented in the comments with their concept, compared to 72.75\%~(243/334) of identifiers of numeric types.

\Conclusion{%
Developer usually documented the variables' identifiers of a numeric data type with their purpose~(69.16\%) and concept~(72.75\%), more than the variables' identifiers of type String which were documented with their purpose~(61.14\%) and concept~(55.46\%).
}

Table~\ref{tab:PCD-Chi-Squire} shows the associations between the groups of identifiers of type String and char vs.~numeric to test the association between these groups. We found that the difference in documenting the concept of an identifier is statistically significant between numeric types and String/char using the Chi-squire test with $p=0.0001$. Numeric types are well documented in terms of their concept as compared to String and char, i.e., numeric is more often conceptualised with meaning. 

Additionally, we found that the information of type directives is strongly related to the data types of numeric, String and char, and boolean variables, with $p<0.05$. For instance, we found that boolean identifiers are not documented in terms of directives~(0/37), as expected, because booleans' range is only two elements. In contrast, developers tend to provide additional information that describes the directives of identifiers of strings and chars more than numeric variables, 30.57\% and 16.17\%, respectively. This is likely because String variables are well directive-documented due to extra information, such as ``nullability'' that is usually documented with the identifiers in the comment.

\begin{table*}\centering
\caption{Percentage of the Variables of Different Data Types Documented with Different Types of Comments in Each Scope}\label{tab:RQ5-result}
\begin{tabular}{l|rrrr|rrrr|rrrr}\hline
\multirow{2}{*}{\textbf{Data Type}} & \multicolumn{4}{c|}{\textbf{Fields}} & \multicolumn{4}{c|}{\textbf{Local variables}} & \multicolumn{4}{c}{\textbf{Parameter variables}} \\
 & \textbf{Inline} & \textbf{Line} & \textbf{Block} & \textbf{Javadoc} & \textbf{Inline} & \textbf{Line} & \textbf{Block} & \textbf{Javadoc} & \textbf{Inline} & \textbf{Line} & \textbf{Block} & \textbf{Javadoc} \\ \hline
boolean & 0.57 & 1.43 & 0.24 & 8.71 & 0.66 & 7.51 & 0.42 & 0.10 & 0.00 & 0.03 & 0.03 & 6.26 \\
byte & 1.04 & 0.48 & 0.04 & 1.40 & 0.17 & 3.28 & 0.16 & 0.02 & 0.00 & 0.01 & 0.02 & 1.55 \\
char & 0.02 & 0.05 & 0.02 & 0.58 & 0.04 & 0.45 & 0.04 & 0.00 & 0.00 & 0.00 & 0.01 & 0.46 \\
int & 1.53 & 2.66 & 0.47 & 25.75 & 2.95 & 28.82 & 1.58 & 0.33 & 0.00 & 0.14 & 0.18 & 15.54 \\
short & 0.02 & 0.03 & 0.02 & 0.92 & 0.04 & 0.27 & 0.00 & 0.00 & 0.00 & 0.00 & 0.00 & 0.22 \\
long & 0.26 & 0.75 & 0.15 & 4.98 & 0.41 & 4.48 & 0.23 & 0.07 & 0.00 & 0.02 & 0.02 & 4.40 \\
float & 0.07 & 0.12 & 0.02 & 1.58 & 0.21 & 1.60 & 0.05 & 0.03 & 0.00 & 0.01 & 0.01 & 1.29 \\
double & 0.38 & 0.26 & 0.06 & 1.92 & 0.82 & 5.84 & 0.63 & 0.05 & 0.00 & 0.04 & 0.02 & 2.72 \\
String & 0.96 & 2.96 & 0.63 & 38.93 & 1.76 & 35.30 & 1.45 & 0.19 & 0.00 & 0.20 & 0.22 & 66.60 \\ \hline
\end{tabular}
\end{table*}

\subsection{RQ3: How frequently are primitive variables documented in comments?}\label{sub:RQ3}

Next, we used our large-scale data set which contains 1,703,234~(19.70\%) commented variables~(i.e., developers wrote a comment above the declaration of the variable) out of a total of 8,646,435 primitive variables~(see Table~\ref{tab:dataset-evaluation-and-large-scale-analysis-haveComment}), to investigate how often developers documented the variables’ identifiers in comments, i.e., developers mentioned the variables’ identifiers in the associated comments. Out of 1,703,234 commented variables, our matching approaches were able to detect 67.40\%~(1,147,947) of the identifiers that were documented in their related comments, using the union approach, as its F-score scored the best value~(0.986) among all detecting approaches~(see RQ1). We used 1,147,947 documented variables to answer RQ3 and subsequent research questions: RQ4 and RQ5. We now report the findings to answer RQ3.

Figure~\ref{fig:documnted-notDocumnted-notCommented-variables} shows for each of the data types 1) the percentage of the variables that have comments and are documented in their accompanying comments, 2) the percentage of the variables that have comments but are undocumented~(i.e., not mentioned in their accompanying comments), and 3) the percentage of variables that do not have comments.

We can see from the Figure that the percentage of the documented variables in the comments compared to undocumented variables is high except for the int data type where the opposite was observed. Moreover, 17.9\% of String variables were documented in their comments compared to undocumented~(5.2\%), which showed the highest ratio among all the documented variables followed by byte, boolean and float. This can be explained, as developers tend to document the identifiers of String more than other data types due to its capability to store any data ranging from textual information to symbols and numbers, which might be required from the developers to provide additional information of type directive to describe the content of the String variables. In addition, 80.30\%~(6,943,201) of primitive variables were found to be uncommented, where the long data type showed the highest ratio of uncommented variables among other primitive types. The String data type, on the other hand, showed the lowest ratio of uncommented variables compared to other primitive variables.

\Conclusion{%
80.30\% of the variables do not have comments, and among those commented variables, 13.28\% of the variables were documented in their comment, while 6.42\% of the variables were undocumented.
}

\subsection{RQ4: What are the distributions of documented variables by their scope of declarations?}
As reported in Section~\ref{sub:RQ3}, 13.29\%~(1,147,947) of the variables were documented in their related comments. For these 1,147,947 variables, Figure~\ref{fig:data-types-per-scopes} investigates the distributions of variables’ data type in their accompanying comments in each scope. Identifiers of String data type declared in methods’ parameters were usually documented~(67.02\%) in their associated comments. In contrast, identifiers of the data types of int and boolean were most often documented in local and field scopes, 33.70\% and 10.95\%, respectively.

Comparing data types of each variable with their declaration scopes, we found that String 63.39\%~(727,631), int 19.12\%~(208,058), and boolean 6.91\%~(79,283) data types were frequently used and documented across all the scopes.

On the other hand, the lowest number of variables documented in the comments were short~(0.31\%), char~(0.49\%), float~(1.39\%) and double~(2.93\%) data types across all the scopes of declaration.

Variables declared in methods’ parameters were more documented~(85.32\%) in their comments than those declared in the scopes of field~(10.98\%) and methods’ bodies~(3.70\%). The high number of documented variables in parameters might be because comments were automatically generated by Javadoc and usually mentioned the variables’ identifiers in the comment. In contrast, the lower numbers of the variables documented in the methods’ bodies were due to their smaller scope, where developers tend not to document these identifiers in the comments.

\Conclusion{%
Across all scopes, developers often use and document variable identifiers of type String~(63.39\%), int~(19.12\%) and boolean~(6.91\%) in their accompanying comments more than other primitive data types. Variables declared in methods’ parameters were more documented in their comments than those declared in scopes of field and methods’ bodies.
}

\subsection{RQ5: What are the types of comments associated with the scopes of the documented variables?} 
Finally, we report our result to reveal the types of comments associated with the documented variables in each scope, as shown in Table~\ref{tab:RQ5-result}. The results can be interpreted in the same way as those in Figure~\ref{fig:data-types-per-scopes}, except that in this case, we show the types of the comments~(e.g., inline, line, block and Javadoc) associated with these data types and variables per scope. Our result reveal that a vast majority of the variables declared in the methods’ parameters were documented in a comment of type Javadoc 99.04\%~(969,976), followed by block 0.5\%~(4,878). Similarly, out of 126,093 variables declared in field scope, we found 84.77\% of these variables were documented using Javadoc, followed by line comment~(8.76\%). Finally, 87.55\% of local variables were documented using line comment, followed by inline comment~(7.08\%).

\Conclusion{%
97.41\% of the variables declared in field and parameter scopes were documented in Javadoc comments, while 87.55\% of local variables were documented using line comments.}

\section{Threats to validity}\label{sec:thrests-validity}

Just like any empirical study, for our study, there are a number of threats that may affect our results’ validity.

First, the taxonomy used to categorise the documentation of variables was based on three types of knowledge inspired by Maalej and Robillard~\cite{maalej2013patterns}, and adapted to fit our purpose. Future work should explore other dimensions of these types of knowledge. 

The number of primitive variables used to evaluate our detection approaches is small, which can introduce bias but is a necessary limitation due to the effort required for manual annotation. Besides, the recall reported in this paper is an approximation, as a theoretical value of the recall cannot be computed precisely because much effort to manually annotate the data would be required.

Finally, our work may not generalise to other programming languages, especially those that are loosely typed. Therefore, we cannot claim that our findings generalise beyond the evolution data set used in this paper.

\section{Related Work}\label{sec:related-work}
In this section, we discuss literature related to approaches for the detection of code comment inconsistencies and taxonomies of knowledge types.

\textbf{Automatic Assessment of Comment Quality}:
Researchers have developed tools and metrics to measure the quality of code comments. For example, Khamis et al.~\cite{khamis2010automatic} developed a tool called JavadocMiner based on Natural Language Processing~(NLP) to assess the quality of Javadoc comments by evaluating the ``quality'' of the language used in the comment as well as its consistency with the source code. The quality of the language is assessed using readability metrics such as the Gunning Fog Index and combined with the several heuristics, e.g., checking whether the comment uses well-formed sentences, including nouns and verbs. Furthermore, the consistency between code and comments is checked with a heuristic-based approach. For example, a method having a return type and parameters is expected to have these elements documented in the Javadoc with @return and @param.   

In another work, Steidl et al.~\cite{steidl2013quality} proposed a machine learning model for comment quality analysis and assessment based on various comment categories including header comments, member comments, in-line comments, section comments, code comments and task comments. The model described the quality attributes in terms of coherence, consistency, completeness, and usefulness, and assessed two quality metrics. The validity of the model was then evaluated with a survey among 16 experienced developers.

Similarly, Sun et al.~\cite{sun2016code} extended the work of Steidl et al.~by evaluating code comments in jdk8.0 and jEdit to provide comprehensive comment assessment and recommendation. The study consists of header comment analysis and method comment analysis that highlights the correlation between method’s name and comment.

Hata et al.~\cite{hata20199} investigate the role of links contained in source code comments. They find that licenses, software homepages, and specifications are among the most prevalent types of link targets, and that links are often used to provide metadata or attribution.

Hao~\cite{he2019understanding} carried out an approach to understand comments in source code, investigating whether projects practice commenting differently and what may cause the differences. They chose five programming languages including JavaScript, Java, C++, Python and Go. They found that the most commented project was a Java design patterns project and that comments in Java and Python were more prevalent than in C++.

Other studies have mainly focused on the automatic detection of code-comment inconsistencies such as the work presented by Tan and colleagues~\cite{tan2007icomment, tan2012tcomment}. In their work, they presented iComment\cite{tan2007icomment} which combined program analysis, a technique using NLP and machine learning to detect code-comment inconsistencies. iComment can detect inconsistencies related to the usage of locking mechanisms in code and their description in comments. They also presented @TCOMMENT in their follow-up work~\cite{tan2012tcomment}. @TCOMMENT is an approach which is able to test the consistency between Javadoc comments related to null values and exceptions with the behaviour of the related method’s body.  

A rule-based approach named Fraco was proposed by Ratol et al.~\cite{ratol2017detecting} to detect code-comment inconsistencies resulting from rename refactoring operations performed on identifiers. Their evaluation shows the superior performance achieved by Fraco as compared to the rename refactoring support implemented in Eclipse. Our work, while not related to the automatic assessment of comments’ quality, provides an empirical study on how developers tend to document identifiers of primitive variables in source code comments.

\textbf{Taxonomy of knowledge types}:
Padioleau et al.~\cite{padioleau2009listening} proposed a taxonomy based on meanings of comments and manually classifiers 1,050 comments. They found that 52.6\% of these comments can be leveraged to improve software reliability and increase developer productivity. Monperrus et al.~\cite{monperrus2012should} empirically studied API directives which are constraints about usages of APIs and they built a corresponding taxonomy. Maalej and Robillard~\cite{maalej2013patterns} leveraged grounded methods and analytical approaches to build a taxonomy of knowledge types in API reference documentation and manually classified 5,574 randomly sampled documentation units to assess the knowledge they contain. Unlike those studies, we developed a taxonomy of knowledge types for identifiers detected in the code comments as the existing taxonomies do not fit the purpose of this paper.

\section{Implications and Future Work}\label{sec:conclusion}

Primitive variable data types are fundamental elements in any programming language. Because of the nature of these variables, which requires developers to inject meaning in the accompanying comments, their role might be difficult to understand if not sufficiently explained in code comments, which can ultimately reduce the productivity of developers.

Our results indicate which variables in what context tend to be documented and which ones are not. Knowing which primitive variables~(i.e., most of the variables in our data set) tend to be documented and what knowledge this documentation contains are first steps towards raising awareness among developers about what should be documented and what is often missing.  This can provide a pathway towards integrated automated tooling to assess the presence of knowledge types and to potentially issue recommend changes in cases of lack of knowledge types in the comment related to the variables' identifiers. 

In future work, we plan similar work on non-primitive variables and consider expanding the taxonomy of knowledge types regarding variables’ identifiers documented in comments to cover more knowledge types, which can help reveal additional information linked to these identifiers.

\IEEEtriggeratref{20}
\bibliographystyle{IEEEtran}
\bibliography{refe}

\end{document}